\documentclass[a4paper]{article}

\usepackage{INTERSPEECH2018}
\usepackage[draft]{hyperref}
\usepackage{booktabs}
\usepackage{subfig}

\title{Acoustic Features Fusion using Attentive Multi-channel Deep Architecture}
\name{Gaurav Bhatt$^1$, Akshita Gupta$^1$, Aditya Arora$^1{}^2$, Balasubramanian Raman$^1$}
\address{
  $^1$Indian Institute of Technology Roorkee
  $^2$ Inception Institute of Artificial Intelligence, Abu Dhabi
  }
\email{gauravbhatt.deeplearn@gmail.com, akshitadvlp@gmail.com, adityadvlp@gmail.com, balarfma@iitr.ac.in}

\begin{document}

\maketitle
\begin{abstract}
In this paper, we present a novel deep fusion architecture for audio classification tasks. The multi-channel model presented is formed using deep convolution layers where different acoustic features are passed through each channel. To enable dissemination of information across the channels, we introduce attention feature maps that aid in the alignment of frames. 
The output of each channel is merged using interaction parameters that non-linearly aggregate the representative features. Finally, we evaluate the performance of the proposed architecture on three benchmark datasets :- DCASE-2016 and LITIS Rouen (acoustic scene recognition), and CHiME-Home (tagging). Our experimental results suggest that the architecture presented outperforms the standard baselines and achieves outstanding performance on the task of acoustic scene recognition and audio tagging.

\end{abstract}
\noindent\textbf{Index Terms}: acoustic scene recognition, audio tagging, deep learning, feature fusion.

\section{Introduction}

While deep architectures have been shown to achieve top performance on simple audio classifications tasks like speech recognition and music genre detection, their application to complex acoustic problems has a significant room for improvements. Two challenging audio classification tasks that have been recently introduced are acoustic scene recognition (ASR) \cite{kong2016deep} and audio tagging \cite{xu2017unsupervised(1)}. ASR is defined as the identification of environments in which an audio is captured, while audio tagging is a multi-label classification task. To solve these challenge, a majority of current research has shown the effectiveness of feature fusion with deep architectures such as deep neural networks (DNN) \cite{kong2016deep}, convolution neural networks (CNN) \cite{Valenti2016(7)}, and recurrent neural models \cite{Bae2016(17),phan2017improved(15)}. A general problem with multi-channel deep networks is the limited memory and low interaction between subsequent layers. To the problem of limited memory, an attention mechanism has been introduced. In this paper, we demonstrate an attention mechanism that can be used to guide the information flow across multiple channels enabling a smoother convergence that results in better performance.
%It has been shown in previous studies that the amalgamation of several acoustic features with deep models improves the performance, however, a concrete reasoning behind this fact is missing from previous works. Also, very few researchers have explored the information flow for deep models when combining acoustic features together. This creates a research gap as an in-depth architectural analysis of learning models is missing from most of these studies, which may be beneficial for several dependent audio tasks

A popular approach for acoustic scene recognition (ASR) and the tagging task is to use the low-level or high-level acoustic features such as Mel-frequency cepstral coefficients (MFCCs), Mel-spectrogram, Mel-bank, log Mel-bank features, etc., with the state-of-the-art deep models \cite{parascandolo2017convolutional(3),phan2017audio(13)}. %Nonetheless, some researchers have explored the complementary properties of acoustic features that aid in the task of classification and tagging \cite{sainath2015learning(4), tu2017information(5)}. 
Some of these acoustic features possess complementary qualities, that is, for two given features, one is apt in identifying certain specific classes, while the other is suitable for the rest. This complementarity property may depend upon the spectrum range in which these features operate. Hence, it is possible to obtain a boost in performance when multiple complementary features are combined together as the overall range over which the learning models can operate is increased. 
For an instance, 
%the wavelet octave coefficients of residues that are generated by applying a pitch-synchronous wavelet transform to the residual signal are shown to have complementary information to the conventional MFCC features. Other examples include 
the augmentation of the delta and acceleration coefficients with MFCC proves to be more effective for acoustic scene classification \cite{Eghbal-Zadeh2016(14),yun2016discriminative}. Similarly, Mel frequency components and the log of Mel components are examples of one such complementary pair that we use in our work. 

In this paper, we combine acoustic features using a multi-channel approach, %Each of the channel is inspired by several state-of-the-art deep models for computer vision such as AlexNet \cite{alexnet(28)}, VGGNet \cite{vgg(29)}, Inception \cite{inception(30)} and Resnet \cite{ressnet(31)} 
where we add subsequent convolution and pooling layers to the input low-level complementary features. 
To effectively amalgamate the properties of several acoustic features we introduce three feature fusion techniques: early fusion, late fusion, and hybrid fusion, depending on the position where acoustic features are fused together (here, position refers to the intermediate neural layers). The early fusion strategy comprises of stacked attention layers that introduce a flow of information between the channels to facilitates better convergence. In the late fusion, we introduce trainable parameters to the model which enables better generalization for the audio classification and tagging tasks. Finally,  we demonstrate the performance evaluation of the proposed model on DCASE-2016 (ASR), LITIS-Rouen (ASR) and CHiME-Home (audio tagging) datasets.

\begin{figure*}
\centering\includegraphics[width=1\textwidth,height=0.35\textheight]{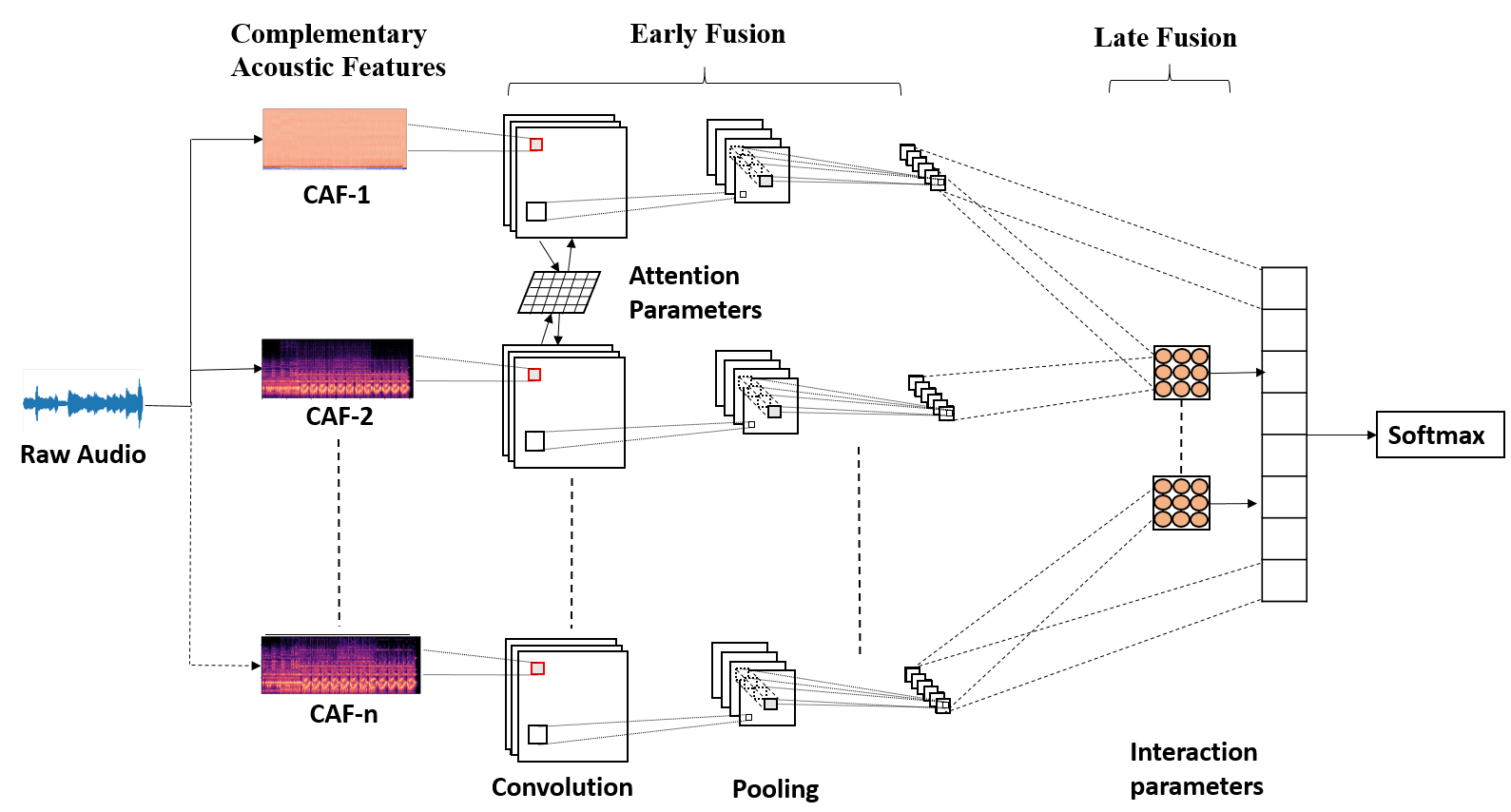}
\caption{The proposed multi-channel deep fusion architecture with complementary acoustic features (CAF) as input.}
\label{fig:one}
\end{figure*}

\section{Related Work}

%We use five type of different feature representation methods:
In this section, we discuss previous work related to audio classification and audio tagging. 
These domains have recently gained popularity because of open challenges such as Dcase2013 \cite{dataset_13(25)}, Dcase2016 \cite{mesaros2016tut}, and Chime2018 \cite{barker2018fifth}. 

For audio classification and tagging, the Mel frequency cepstrum coefficient (MFCC) and the Guassian mixture model (GMM) are widely used as a baseline \cite{Heittola2016(9), Foster2016}. Most published works in this domain uses Mel-spectrograms as features with deep parallel convolution architectures \cite{Valenti2016(7),parascandolo2017convolutional(3), xu2017convolutional}. Traditional techniques focus on using hand-crafted audio features as the input to various machine learning classifiers. Some of the recent research has been focused on passing the term-frequency representation of the waveform through convolution neural networks \cite{Lidy2016(8),xu2017convolutional,parascandolo2017convolutional(3),phan2017improved(15),Valenti2016(7)}, or deep neural networks  \cite{Kong2016,Xu2016,kong2016deep} However, deep networks have not yet outperformed feature-based approaches.

Currently, the main challenges to audio tagging datasets are the uneven distribution of samples, along with uneven labels \cite{foster2015chime}. For audio tagging tasks, attention models have been introduced, which have shown more accurate results than other hybrid combinations of deep models \cite{xu2018large}.

\section{Complementary Acoustic Features}

%We use five type of different feature representation methods:

The Mel and log-Mel are a set of complementary acoustic features (CAF). The Mel frequencies capture classes which lie in the higher frequency domain and log-Mel frequencies capture classes that lie in the lower frequency domain. We conjecture that passing the features via a multi-channel model it is possible to efficiently combine the complementary properties inhibited by these features. We calculate the Mel spectrum by taking the spectrogram of the raw audio and combining it with the transposition of the spectrogram. The Mel frequencies are kept at 40, resulting in 40-dim Mel features with non-overlapping frames and 50\% hop size. 

We also compute the Mel-Frequency components (MFCCs) and select 13 mel frequency ceptral coefficients (including the $0^{th}$ order coefficient) with a window size of 1024 frames with 50\% overlapping. With the time-varying information, we combine the first and second derivatives (i.e. the delta and acceleration coefficients).

For the Constant Q Transformation (CQT) features, we select 80 bins per octave with 50\% hop size.

\section{Multi-Channel Deep Fusion}

In the model presented, each complementary acoustic feature is passed on to a separate CNN, thereby forming a multiple channel architecture. Each channel is formed using 128 kernels in the first layer, with a receptive filter of size 3$\times$3. This gives us the convolved features which are then sub-sampled using a max/global pooling with filter size 2$\times$2. In the second convolution layer, we use a large number of kernels (256) for exploring higher-level representations. The activation function that we use is the rectified linear units (Relu) in the subsequent convolution layers. All the parameters are shared across the layers.

%One  problem  with  long  audio  recordings  is  noisy  channels, as well as  fewer  foreground  events.   To  improve  the  performance of the underlying deep architectures, we follow the works of [5, 3] and divide the audio features into segments. Instead of  using  the  whole  audio  feature  as  the  training  sample,  we break the snippet into T segments of \textit{length = 1024} frames with a hop size of \textit{512 frames}.  This is done to ensure that the underlying learning model is able to capture the important foreground events in the long recordings.

Some problems related to long audio recordings are that the channels can be noisy and the number of foreground events may not be sufficient. To improve the performance of the underlying deep architectures, we follow the work of [5, 3] and divide of the audio features into segments. In place of using the  whole audio feature as the training sample, we decompose the snippet into T segments having \textit{length = 1024} frames with a hop size of \textit{512 frames}.  This is done to ensure that the underlying learning model is able to capture the important foreground events in the long recordings.

\begin{figure}[b!]
\centering\includegraphics[width=0.4\textwidth,height=0.2\textheight]{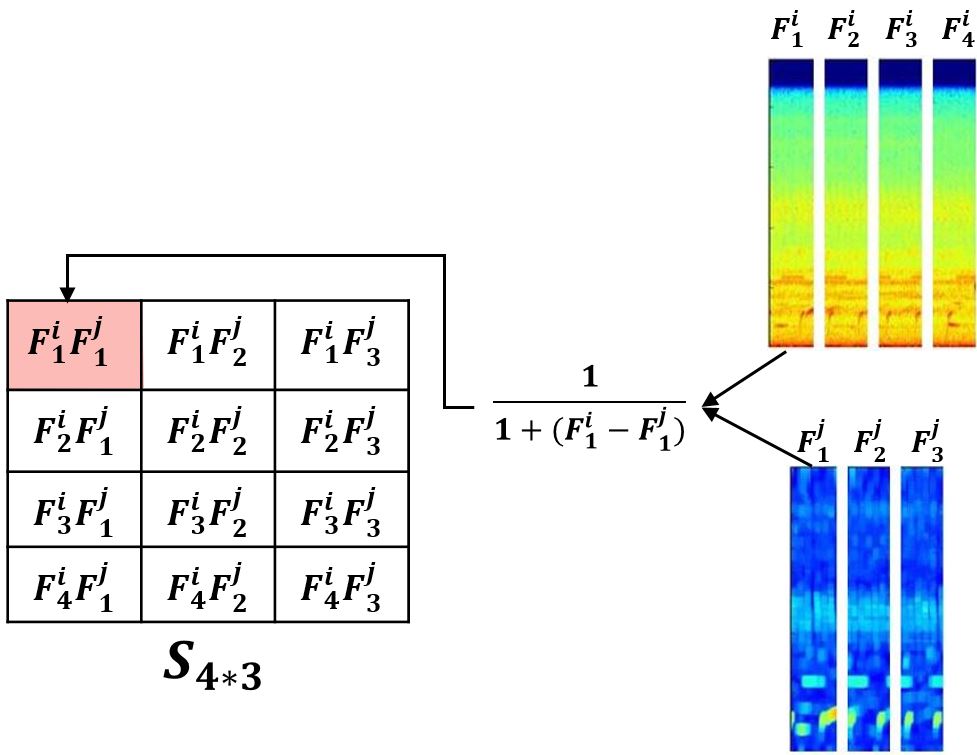}
\caption{Computing similarity matrix $S_{4*3}$ for two acoustic features, $F^i$ and $F^j$, with frame sizes of 4 and 3 respectively.}
\label{fig:one}
\end{figure}

\subsection{Early fusion}
%To ameliorate the performance of the proposed architecture, we introduce the combination of features within the learning model. 
In the multi-channel architecture, each layer of the channel $i$ captures higher level representation based on the acoustic feature passed through. These representative features differ from the corresponding layer of channel $j$, where $j\ \in \{N-i\}$ (shown in Figure 1). The temporal sequences input can be aligned together with the help of the attention mechanism \cite{yin2015abcnn}, so that the properties of the audio sequences important in one channel can be reflected in the others. We compute an attention matrix to align two audio representative feature maps, which is followed by the addition of trainable parameters to transform the matrix into convolution feature maps (Equation 3 and 4). This is essentially attentive convolution and helps the model to assign a higher score important events than the rest (shown in Figure 2).%Similarly, we compute the similarity matrices in higher convolution layers and use then for pooling. 
%Given the input features

Inspired by \cite{yin2015abcnn}, we introduce a similarity feature matrix $S_{ij}$ that can influence convolutions across multiple channels, where $S_{ij}$ is shared across channels $i$ and $j$. The similarity features assign a higher score to those frames in channel $c_i$ that are relevant to frames in channel $c_j$, where $j\ \in \{N-i\}$. 
%Each attention matrix $A_{ij}$ is created by aligning representation features given by channel $i$ to the representation features by channel $j$. 
The values in the row of $S_{ij}$ denote the distribution of the similarity weights of the $t^{th}$ frame of $c_i$ with respect to $c_j$, and the columns of $S_{ij}$ represents the distribution of similarity weights of the $T^{th}$ frame of $c_j$ with respect to $c_i$. As shown in Figure 2, the rows of the similarity matrix $S_{4*3}$ represent the distribution of audio frames of feature $i$ and columns represent the distribution of feature $j$.

Let the representative feature for channel $i$ and $j$ be given by $F_i^r$ and $F_j^r$ respectively.The similarity feature matrix $S_{ij}$ is then computed as 
\begin{align}
S_{ij}^{x,y} &= similarity_{score}(F_i^r[:,x],F_j^r[:,y])
\end{align}
where $similarity_{score}(a,b)$ is given by 
\begin{align}
similarity_{score}(a,b) &= \frac{1}{1+|a-b|}
\end{align}
Using the similarity matrix $S_{ij}$, the attentive feature representations $F_{k\in\{i,j\}}^s$ are computed as 
\begin{align}
F_{i}^s &= W_i \cdot S_{ij}^T\\
F_{j}^s &= W_j \cdot S_{ij}
\end{align}

The channel representative features $F_i^r$ and the attentive representative features $F_{i}^s$ are stacked together as an order-3 tensor matrix before passing to further convolution layers.
%The attention feature maps are 

The idea of attentive convolution works in the early stages and we call it early fusion technique.

\subsection{Late Fusion}
Once all the features maps are obtained using the multi-channel architecture, we use an interaction matrix $W$ to compute the interaction between them (shown as interaction parameters in Figure 1). Given the feature maps for channel $i$ and $j$ are $F_{i}^p$, $F_{j}^p$ respectively, the interaction score is computed as
\begin{align}
Score_I(F_{i}^p,F_{j}^p) = (F_{i}^p)^T\cdot W \cdot F_{j}^p
\end{align}
We share a common weight matrix $W$ across all the possible pairs. This is done to ensure that all the channels interact with each other and the computation burden over the network is reduced.

\subsection{Parameters Sharing}

The proposed multi-channel architecture, along with early and late fusion, increases the number of trainable parameters, which results in optimization issues. We solve this problem by sharing the parameters across layer $i$. That is, the convolution layer shown in Figure 1 is shared across all the channels. We also share the attention weights $W_i$ and $W_j$, as computed by Equation 3 and 4. The interaction matrix $W$ used in the late fusion is also shared across all the channels. Finally, we combine the representative features computed at each pooling layer with the similarity scores (shown in Figure 1 as late fusion). This ensures that the model can capture representations at each level of abstraction.

\begin{table}[t!]
\caption{Hyper-parameters used in feature fusion layer for training.}
\bigskip
\centering
\begin{tabular} {|c|c|}
\hline
\textbf{Hyper-parameter} & \textbf{Value}\\
\hline
 Depth & 6 \\
 Number of neurons & 600\\  
 Regularization & L2 ; dropout \\
 L2 & 0.005 \\
 Dropout & 0.3\\
 Optimizer & adam\\
 Loss & Binary crossentropy\\
 Learning rate & 0.001\\
 Batch size & 100\\
 Number of Epochs & 10\\
 filter Length & (60,3)\\
 Number of filters & 256\\
\hline
\end{tabular}
\end{table}

\begin{table*}
\caption{Performance of various architectures on ASR and Audio Tagging. The first half of each table shows the baseline results, the middle section is the performance of state-of-the-art systems and the bottom section is the performance of the proposed model.}
\bigskip
\centering
\subfloat[DCASE-2016]{%
\scalebox{0.97}{
\begin{tabular} {|c|c|}
\hline
\hline
  \textbf{Techniques} & \textbf{Accuracy \%} \\
\hline
MFCC + GMM & 72.50 \\
Mel + DNN \cite{kong2016deep} & 81.00 \\  
 \hline
MFCC + fusion \cite{Eghbal-Zadeh2016(14)} & \textbf{89.70} \\
Spectrogram + NMF \cite{Bisot2016(16)} & 87.70 \\
% Various + fusion \cite{park2016score(10)} & 87.20 \\

 %\hline
 LTE-fusion + CNN \cite{phan2017improved(15)} & 81.20 \\
 BiMFCC + I-vector \cite{li2017comparison} & 81.70 \\
 Mel + CNN \cite{Valenti2016(7)}& 86.20 \\
% BiMFCC + DNN \cite{li2017comparison}  & 84.20 \\
BiMFCC + RNN \cite{li2017comparison} & 80.20 \\
BiMFCC + Fusion  \cite{li2017comparison}& 88.10 \\
 \hline

 Proposed & 88.70 \\
\hline
\hline
\end{tabular}}}%
\qquad% --- set horizontal distance between tables here
\subfloat[LITIS-Rouen]{%
\scalebox{0.97}{
\begin{tabular} {|c|c|c|c|}
\hline
\hline
  \textbf{Techniques} & \textbf{P \%} & \textbf{F1 \%} \\
\hline
%HOG + CQT \\
% HOG + CQT & 91.50 &- \\
DNN + MFCC & 92.20 &- \\
 \hline
% \iffalse
CNN+Fusion \cite{phan2017improved(15)} & 96.30 & 96.50 \\
RNN+Fusion \cite{phan2017audio(13)} &97.50 & 97.70  \\
LTE+Fusion \cite{phan2016label} &95.90 &96.20  \\
% Kernel PCA \cite{bisot2016acoustic} &- & 95.60 \\
% Sparse NMF \cite{bisot2016acoustic} &- & 94.10\\
% Conv. NMF \cite{bisot2016acoustic} &- & 94.50 \\
 \hline
Proposed & \textbf{98.00} & \textbf{98.25} \\
\hline
\hline
%\fi
\end{tabular}}}
\qquad
\subfloat[CHiME-Home]{%
\scalebox{0.97}{
\begin{tabular} {|c|c|}
\hline
\hline
  \textbf{Techniques} & \textbf{EER} \\
\hline

MFCC + GMM \cite{Foster2016} & 21.0 \\
Mel + DNN \cite{Kong2016}& 20.9 \\
  
\hline
CQT + CNN \cite{Lidy2016(8)} & 16.6 \\
MFCC + GMM \cite{yun2016discriminative} & 17.4 \\
MFCC + DNN \cite{Xu2016} & 17.85 \\
DAE + DNN \cite{xu2017unsupervised(1)} & 14.8 \\
Mel + IMD \cite{xu2017convolutional} & \textbf{12.3} \\

log-mel + CRNN \cite{parascandolo2017convolutional(3)} & \textbf{11.3$\pm$0.6} \\
 \hline
Proposed& \textbf{14.0} \\
\hline
\hline
\end{tabular}}}%

\end{table*}

\section{Experiments and Results}

\subsection{Dataset Description}
% For experimentation, we use the following two benchmark datasets: DCASE-2016 (task 1) and LITIS Rouen Dataset. 

\textbf{DCASE-2016}. %In this dataset, each audio recording (originally 3-5 minute duration) is divided into 30-second segments. The acoustic scenes are comprised of indoor scenes like cafe/restaurant, grocery store, home, library, metro station, and office. The outdoors scenes include bus, city center, forest path, lakeside beach, residential area, and urban park. Finally, the vehicle/traveling scenes are categorized as train and trams.
The dataset consists of 1560 samples which are divided into the development dataset (1170) and evaluation dataset (390). Each audio class in the development set consists of 78 samples (39 minutes of audio) while the evaluation dataset is comprised of 26 samples (13 minutes of audio) for each class. The organizers of DCASE-2016 have provided the four cross-fold validation meta-data which is used for tuning the parameters of the network.

\textbf{LITIS Rouen}. This dataset consists of audio recordings of 30 second duration, which are divided into 19 classes. The total number of audio samples is 3026. %The distribution of samples in respective classes is uneven with \textit{Avion} class having only 23 audio samples.
We divide the dataset into 10 cross-validation sets with 80:20 random splits each. The final performance of the proposed techniques is computed by averaging the accuracy on all 10 test sets.

\textbf{CHiME-Home};. This dataset consists of audio recordings of 4-second duration in two sampling frequencies: 48Khz in stereo and 16Khz in mono.  We use mono audio data with the 16Khz sampling frequency, which is further divided into 7 classes. The total number of audio samples is 2792, which are divided into 1946 development sets and 846 evaluation sets. %The distribution of samples in respective classes is uneven with \textit{Avion} class having only 23 audio samples.
Each piece is annotated with a single or multiple labels. \cite{parascandolo2017convolutional(3)}.

\iffalse
\subsection{Dataset Description}
%For experimentation, we use the following two benchmark datasets: DCASE-2016 (task 1) and LITIS Rouen Dataset. 

\textbf{LITIS Rouen}. This dataset consists of audio recordings with 30 seconds duration which is divided into 19 classes. The total number of audio samples is 3026. %The distribution of samples in respective classes is uneven with \textit{Avion} class having only 23 audio samples.
We divide the dataset into 10 cross-validation sets with 80:20 random split each. The final performance of the proposed techniques is computed by averaging the accuracy on all 10 test sets. For this task, the precision and recall are chosen as the evaluation metrics.

\textbf{CHiME-Home}. This dataset consists of audio recordings of 4 seconds duration in two sampling frequencies 48Khz in stereo and 16Khz in mono. We, therefore, use mono audio data with the 16Khz sampling frequency which is further divided into 7 classes. The total number of audio samples is 2792 which is divided into 1946 development set and 846 evaluation set. %The distribution of samples in respective classes is uneven with \textit{Avion} class having only 23 audio samples.
Each chunk is annotated with single or multi-labels \cite{parascandolo2017convolutional(3)}. For this task, the equal error rate (EER) is chosen as the evaluation metric.
\fi

\subsection{Hyperparameters}

The architecture details of the individual channels are described in Table 1. The output of the global average pooling layers is concatenated and then passed  on  to  the  intermediate matrix,  which computes the interaction between them. Finally, we use \textit{adam} as the optimizer for binary cross-entropy loss.

\subsection{Baselines}
For DCASE-2016 we use the Gaussian Mixture Model (GMM) with MFCC (including acceleration and delta coefficients) as the baseline system. This baseline is provided by DCASE-2016 organizers. The other baseline used is DNN with mel-components \cite{kong2016deep}. For LITIS-Rouen we use the HOG+CQA and DNN + MFCC results as the baseline. These results are taken from \cite{phan2017audio(13)}. For the CHiME-Home dataset, we use the standard baseline of the MFCC+GMM system \cite{Foster2016} and mel+DNN \cite{Kong2016}.

\begin{table}
\caption{Ablation results of the proposed model. Eight system configurations for the proposed deep fusion architecture.}
\bigskip
\centering
\scalebox{1}{
\begin{tabular} {|c|c|c|c|}
\hline
\hline
  \textbf{Techniques} & \textbf{DCASE} & \textbf{ROUEN (F1)} & \textbf{CHIME} \\
\hline
Vanilla & 85.50 &96.85 &15.6\\
\hline
EF & 86.1 &96.36 &15.0\\
 LF & 87.00 & 96.80 & 14.6\\
 EF+LF& \textbf{88.70} & \textbf{98.25} & \textbf{14.0}\\
\hline
\hline
\end{tabular}
}
\end{table}

\subsection{Results}

The results for the task of ASR on DCASE-2016 and LITIS-Rouen dataset are shown in Table 2 (a) and (b), respectively. The use of deep fusion achieves the highest precision and F1 measure when compared to current state-of-the-art techniques on LITIS Rouen, while we achieve an accuracy of \textbf{88.7} on DCASE-2016, which is comparable to current top methods.. A similar technique of feature fusion, CNN+Fusion, was used by \cite{phan2017improved(15)}, where they used label-based embeddings. However, our proposed method of fusing complementary features results in better performance than all of their architectures. 

For the task of audio tagging (Table 2 (c)), the proposed model achieved an equal error rate (EER) of \textbf{14.0}, which is better than the baselines and all DCASE-2016 challenge submissions. These submissions all omitted the \textit{silence} class. In contrast, we keep the \textit{silence} class and omit the \textit{others} class,  since it has high variance due to the introduction of random samples. We keep the \textit{silence} class instead. %The performance of the proposed architecture is considerably higher when compared to the audio tagging task in DCASE-2016 challenge submissions \cite{parascandolo2017convolutional(3),xu2017convolutional}. 
In the task of audio tagging, the temporal models outperformed the non-temporal techniques and the reason being the dependence of temporal sequences on multi-label classification. As shown in Table 2 (b), most of the successful systems consist of CNN and RNN based models.

Finally, the ablation results are presented in Table 3, which demonstrates the performance of the deep fusion techniques introduced in our work. Here, Vanilla represents the basic feature fusion model without attention and interaction matrix. Here, we present four systems describing the in-depth analysis of the proposed architecture - vanilla (no fusion), early fusion (EF), late fusion (LF) and hybrid (EF+LF).

\subsection{Discussion}

\begin{figure}
\centering\includegraphics[width=0.48\textwidth,height=0.28\textheight]{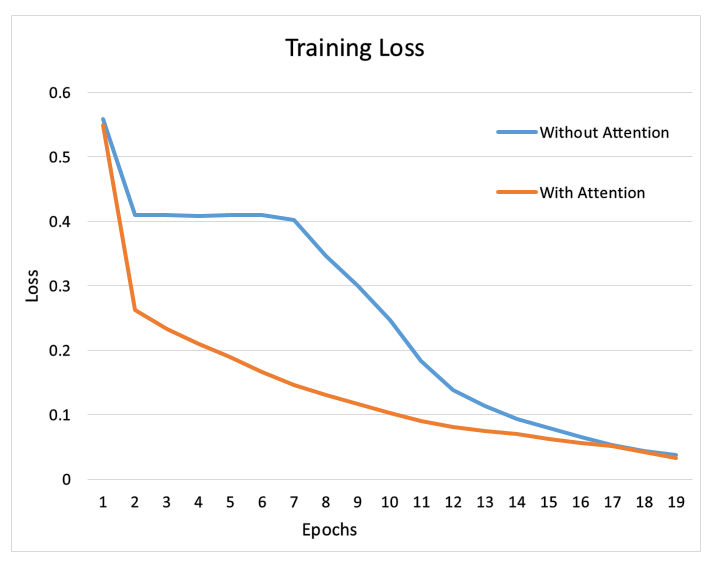}
\caption{Training loss curves for the vanilla system along with the proposed attentive network.}
\label{fig:one}
\end{figure}

 The model combining early and late fusion (EF+EL) had the highest performance amongst all the introduced models. This is due to the enhanced interactivity across the channels. As shown in Table 3, the vanilla system is an amalgamation of multiple features without any fusion and acts as a baseline model. The early and late fusion models achieve a better performance than the baseline model.  The LF system constitutes the fusion-by-multi-channel architecture, where the interaction parameters are responsible for non-linear feature augmentation. The similarity features are accompanied by additional trainable parameters, which result in higher performance but are computationally expensive to train.

Finally, we present the training curves for the vanilla model and the proposed architecture (Figure 3 and 4). We keep track of the mean square error (MSE) for each iteration along with the binary cross-entropy loss. This is done for the vanilla model and the proposed attention-based model. The training loss and MSE for the attention-based systems show a steep decrease in  loss as compared to the vanilla model. Not only are the decreases in loss quick but the overall losses for attention-based models are also lower than that of the vanilla system. This demonstrates that the introduced attention and similarity parameters are responsible for a smoother convergence. 

\begin{figure}
\centering\includegraphics[width=0.48\textwidth,height=0.28\textheight]{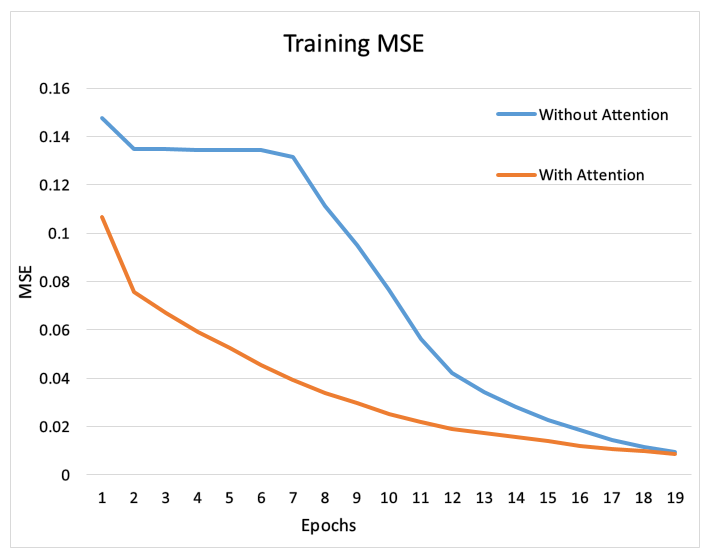}
\caption{Training MSE curves for the vanilla system along with the proposed attentive network.}
\label{fig:one}
\end{figure}

\section{Conclusion}
In this paper, we present a multi-channel architecture for the fusion of complementary acoustic features. Our idea is based on the fact that the introduction of attention parameters between the channels results in better convergence. The proposed technique is general and can be applied to any audio classification task. A possible extension to our work would be to use the pairs or triplets of audio samples of similar classes and pass them through the multi-channel architecture. This could help to align the diverse audio samples of similar classes, making the model robust to audio samples that are difficult to classify.

%From Table 1 it is evident that the proposed techniques achieve state-of-the-art performance on the task of ASR and audio tagging. The ablations results are shown in Table 1 (c) suggests that the introduction of early and late fusion systems show an increase in performance. %The presented idea in this work is rather general to audio classification and can be applied to complex audio-related task as well.

\section{Acknowledgement}
We would like to express our gratitude towards Institute Computer Center
(ICC), and the Indian Institute of Technology Roorkee, for providing us with the necessary resources for this work.

\bibliographystyle{IEEEtran}

\bibliography{mybib}

\end{document}